\documentclass[aps,prl,groupedaddress,showpacs,floatfix,twocolumn]{revtex4-1}

\usepackage{graphicx,graphics,epsfig,subfigure,times,bm,bbm,amssymb,amsmath,amsfonts,mathrsfs}
\usepackage{amsthm}
\usepackage[pdfstartview=FitH]{hyperref}
\usepackage[pdftex]{color}
\usepackage{MnSymbol,wasysym}

\makeatletter
\@ifundefined{textcolor}{}
{%
 \definecolor{BLACK}{gray}{0}
 \definecolor{WHITE}{gray}{1}
 \definecolor{RED}{rgb}{1,0,0}
 \definecolor{GREEN}{rgb}{0,1,0}
 \definecolor{BLUE}{rgb}{0,0,1}
 \definecolor{CYAN}{cmyk}{1,0,0,0}
 \definecolor{MAGENTA}{cmyk}{0,1,0,0}
 \definecolor{YELLOW}{cmyk}{0,0,1,0}
}

\newcommand{\ket}[1]{\left\vert#1\right\rangle}
\newcommand{\bra}[1]{\left\langle#1\right\vert}

\newcommand{\eq}[1]{Eq.(\ref{#1})}

\newcommand{\ignore}[1]{}

\newcommand{\beq}{\begin{equation}}
\newcommand{\eneq}{\end{equation}}
\newcommand{\beqnn}{\begin{equation*}}
\newcommand{\eneqnn}{\end{equation*}}
\newcommand{\beqy}{\begin{eqnarray}}
\newcommand{\eneqy}{\end{eqnarray}}
\newcommand{\beqynn}{\begin{eqnarray*}}
\newcommand{\eneqynn}{\end{eqnarray*}}
\usepackage{graphicx}

\begin{document}

\title[Fundamental filter functions]
%{Fundamental Filter Functions for Dynamical Noise Filtering in Open Quantum Systems}
{A General Transfer-Function Approach to Noise Filtering in Open-Loop Quantum Control}

\author{Gerardo A. Paz-Silva and Lorenza Viola} 

\affiliation{ \mbox{Department of Physics and Astronomy, Dartmouth College,
6127 Wilder Laboratory, Hanover, New Hampshire, USA}}

\date{\today}

\begin{abstract}
We present a general transfer-function approach to noise filtering in open-loop Hamiltonian engineering 
protocols for open quantum systems.  We show how to identify a computationally tractable set of 
{\em fundamental filter functions}, out of which arbitrary transfer filter functions may be assembled 
up to arbitrary high order in principle.  Besides avoiding the infinite recursive hierarchy of filter functions 
that arises in general control scenarios, 
this fundamental filter-functions set suffices to characterize the error suppression capabilities of the 
control protocol in both the time and frequency domain.  We prove that the resulting notion of {\em 
filtering order} reveals conceptually distinct, albeit complementary, features of the controlled dynamics 
as compared to the order of error cancellation, traditionally defined in the Magnus sense. 
Examples and implications are discussed.  
\end{abstract}


\maketitle

Hamiltonian engineering via open-loop quantum control provides a versatile and experimentally  
validated framework for manipulating the dynamics of a broad class of open quantum systems 
\cite{QECBook}. Applications range from dynamical decoupling (DD), 
composite pulse sequences and dynamically corrected quantum gates (DCGs), to 
noise spectroscopy and quantum simulation -- see e.g. 
\cite{dd,KhodjastehLidar:04,Uhrig:07,KenReview,dcgs,spectroscopy,simulation} 
for recent contributions. In this context, {\em generalized transfer filter function} (FF) 
techniques motivated by control engineering are providing an increasingly important 
tool for understanding the dynamical response of the target system in Fourier space 
and for quantitatively analyzing the control performance
~\cite{Kofman:01,PhysRevB.77.174509,WDD,Todd}. 
In particular, this formalism has proved remarkably successful in predicting operational fidelities 
for a variety of control settings in recent trapped-ion experiments \cite{Soare2014}, as long as noise is 
sufficiently weak for low-order approximations to be viable. 

From a control-theory standpoint, a filtering approach to dynamical error 
suppression is desirable for a variety of reasons.
Besides providing an open-loop counterpart to the transfer-function 
perspective that is central to both classical and quantum feedback networks \cite{networks}, 
FF techniques allow, when available, for a substantially more efficient
analysis of the underlying noisy dynamics than direct simulation \cite{Chingiz,Witzel2013}.
Furthermore, unlike traditional time-dependent perturbative 
approaches (such as the Magnus expansion \cite{QECBook,Blanes:08}), 
a frequency-space picture may open up new possibilities for tailoring control synthesis 
and optimization to {\em specific} spectral features of the noise. 
The existing FF framework suffers, however, from severe limitations. 
Even if formal expressions for gate fidelity may be given based on an 
{\em infinite recursive hierarchy} of generalized FFs \cite{Todd}, higher-order FFs 
become rapidly intractable.  Thus, explicit calculations have largely focused thus far on 
single-qubit controlled dynamics in the presence of {\em classical} noise, by truncating 
this recursion to the lowest order and additionally exploiting {\em Gaussian} noise statistics. 
As experiments are revealing the importance of higher-order terms outside the 
weak-noise regime \cite{Soare2014}, and are rapidly progressing toward coupled-qubit systems 
and/or more complex, non-Gaussian error models \cite{Oliver2}, overcoming these limitations 
becomes imperative for further progress.  An additional shortcoming arises from the 
fact that only in very special situations where a {\em single} FF suffices to characterize the 
controlled dynamics, one may unambiguously identify the ``order of error suppression'' of the 
protocol, associated to the order of cancellation in the Magnus expansion, with 
its  ``filtering order'', determined by the low-frequency behavior of the FF \cite{Uhrig:07,Todd}.  
Thus, the present theory does not lend itself to the identification of relevant filtering notions for 
general control scenarios. 

In this Letter, we introduce a filtering framework for open-loop dynamical control based on an  
easily computable set of {\em fundamental filter functions} (FFFs). FFFs replace the unwieldy 
recursive construction of existing schemes by  allowing for arbitrary high-order FFs to be expressed 
as simple linear combinations of products of elements in the set.  We show that FFFs encode all the 
relevant information about the error-suppressing capabilities of a control protocol in both the 
time and frequency domain.  In addition, they suffice to rigorously characterize the (minimum) order of 
{\em cancellation} vs. {\em filtering} that can be guaranteed when no specific assumptions 
are made on the noise model -- other than the bath operators be norm-bounded, 
in the same spirit of DCG theory \cite{dcgs} and non-Markovian quantum error correction 
\cite{Preskill2013}.  We prove that cancellation and filtering are {\em inequivalent notions}, 
with high-order cancellation in the Magnus sense not implying high-order filtering in general, 
and with both notions being a priori equally significant for assessing the control performance. 
Our results provide a firm foundation for recent analyses where this inequivalence has  
manifested in the context of composite-pulse and Walsh-modulated protocols 
\cite{Chingiz,Soare2014}, as well as a new perspective on dynamical error control strategies, with 
potential implications for quantum fault tolerance.

{\em Control-theoretic setting.--} We consider a general finite-dimensional open quantum system 
$S$ coupled to an uncontrollable environment (bath) 
$B$, whose free evolution is described by a joint Hamiltonian of the form 
$ H(t) = H_S + H_{SB}(t)$, with respect to the interaction picture associated to the physical 
bath Hamiltonian $H_B$.
Open-loop control is introduced via a time-dependent Hamiltonian $H_{\text{ctrl}}(t)$ acting on $S$ 
alone, with the controlled dynamics being represented in terms of an {\em intended plus 
error component}, namely, $H(t)+H_{\text{ctrl}}(t) \equiv H_0(t) + H_e(t)$, and evolution under 
$ H_0(t)$ over time $T$ yielding the desired unitary gate (say, $Q$) on $S$.  In addition to 
the system-bath coupling $H_{SB}(t)$ and non-Markovian control errors  $H_{\text{ctrl,e}}(t)$,  
the error Hamiltonian $H_e(t)$ may or may not include $H_S$, depending on whether the  
objective is noise-suppressed storage (DD, in which case $Q= {\mathbb I}_S$) or a non-trivial logic gate 
(such as in DCGs or composite pulses) \cite{dcgs}.

The effect of $H_e(t)$ may be isolated  by expressing the total (interaction-picture) propagator for 
evolution over time $T$ as $U(T)= U_0(T) \tilde{U}_e(T)
\equiv Q \,\tilde{U}_e(T)$, where the ``error propagator'' $\tilde{U}_e (t) ={\mathcal T}
\exp[-i \int_0^t \tilde{H}_e(s) ds ]$ describes evolution in the toggling-frame defined by the  
ideal Hamiltonian $H_0(t)$ (e.g., in a drift-less, ideal-control setting, $H_S=0=H_\text{ctrl,e}(t)$, hence 
$H_0(t) = H_\text{ctrl}(t)$, $\tilde{H}_e(t) = U_\text{ctrl}(t)^\dagger H_e (t) U_\text{ctrl}(t))$.
Without loss of generality, we may write $H_e(t) = \sum_{u} O_u \otimes B_u(t)$, with respect to Hermitian traceless operators $\{O_u\}$ which, together with ${\mathbb I}_S$, form an operator basis on $S$.
The limit of a classical noise source may be recovered by letting 
$B(t) = \beta (t) \mathbb{I}_B$, with $\beta(t)$ being a classical stochastic process.  
Accordingly, 
\beq
\label{yt}
\tilde{H}_e (t) = U_0(t)^\dagger H_e(t) U_0 (t) \equiv \sum_{u ,v} y_{uv} (t) \, O_v \otimes B_u(t) ,
\eneq
where the (target-dependent) ``control matrix'' $y_{uv}(t) $ encapsulates 
the intended effect of the control.  
Dynamical error suppression aims to perturbatively minimize the sensitivity of the controlled 
evolution to $H_e(t)$, by making $\tilde{U}_e(T)$ as close as possible to a ``pure-bath'' evolution. 
Specifically, 
\beqy
\label{Magnus} 
\tilde{U}_e(T) = e^{ \sum_{\alpha=1}^\infty \Omega_\alpha (T)} 
\equiv e^{-i T (H_{SB}^\textrm{eff}(T) + H_{B}^\textrm{eff} (T) ) } 
\eneqy
can be obtained via a Magnus series expansion \cite{Blanes:08}, where 
the $\alpha$-th order Magnus term $\Omega_\alpha (T)$ is a time-ordered 
integral of a function of $\alpha$-th order nested commutators of $\tilde{H}_e (t_j)$, 
$j\in \{1, \ldots, \alpha\}$, over the volume 
$V_\alpha \equiv \{0\leq t_\alpha \leq t_{\alpha-1}\leq \ldots \leq t_2\leq t_1\leq T\},$
and the second equality in Eq. (\ref{Magnus}) defines the 
relevant effective Hamiltonian and (Hermitian) ``error action'' operator 
$\Omega_e(T) \equiv H^{\text{eff}}(T) T =  i \sum_\alpha \Omega_\alpha(T)$ 
\cite{dcgs}.  

A control protocol $y_{uv}(t)$ is said to achieve {\em cancellation order} 
(CO) $\delta \geq 1$ if the norm of $\Omega_e(T)$ (up to pure-bath terms) 
is reduced, so that the leading-order correction 
mixing $S$ and $B$ scales as $\vert\vert T H_{SB}^\text{eff}(T) \vert \vert ={\mathcal O}(T^{\delta+1})$.
The CO coincides with the {\em decoupling order} in the paradigmatic DD case  
\cite{dd,KhodjastehLidar:04,Uhrig:07,ng-2011}.

{\em Generalized filter functions.--} 
In their most basic form, generalized FFs may be defined by seeking a representation of the control 
action in frequency space directly at the effective Hamiltonian level of Eq. (\ref{Magnus}).
By expressing each of the $\tilde{H}_e (t_j)$ in the $\alpha$-th Magnus term $\Omega_\alpha (T)$ 
using Eq. (\ref{yt}), followed by appropriate index relabeling, the latter reads
\begin{align*} 
\Omega_\alpha (T) & = \hspace*{-1mm}\sum_{\vec{u}, \vec{v}}\hspace*{-0.5mm}
 \int_{V_\alpha} \hspace*{-2mm} d^\alpha \vec{t} \, \sum_{p \in \Pi[\{t_j\}]} 
f^{(p)} (\{ y_{[\alpha] } \} ) 
O_{v_1} \cdots \,O_{v_\alpha}  \\
& \,\,\,\,\otimes B_{u_1} (p(t_1)) \cdots \,B_{u_\alpha} (p(t_\alpha)), 
\end{align*}
\noindent 
where $\vec{u}\equiv (u_1, \ldots, u_\alpha)$, $\Pi[\{t_j\}]$ denotes the set of permutations of the labels $\{t_j\}$,
and the function $f^{(p)}$ depends on products of $\alpha$ control matrix elements,
$y_{[\alpha]} \equiv y_{u_1 v_1}(p(t_1)) \cdots y_{u_\alpha v_\alpha}(p(t_\alpha))$.   Upon 
writing the bath variables in terms of their frequency-Fourier transform, 
$B_u(t) \equiv \int_{-\infty}^\infty  (d \omega/2\pi) e^{i \omega t} {B}_u (\omega)$, 
and assuming sufficient regularity for the relevant integrals to converge, 
the $\alpha$-th order Magnus term takes then the desired form:
\begin{align}
\label{GFF}
\Omega_\alpha(T) &
= { -i}\sum_{\vec{u},\vec{v}}\int \frac{d^\alpha \vec{\omega}} 
{(2\pi)^\alpha} \, G^{(\alpha)}_{ \vec{u} \vec{v}} (\vec{\omega}, T) \\ &  \times  
O_{v_1}  \cdots \,O_{v_\alpha}  \otimes B_{u_1} (\omega_1) \cdots \,B_{u_\alpha} (\omega_\alpha),
\nonumber
\end{align}
where the {\em $\alpha$-th order generalized FF $G^{(\alpha)}_{\vec{u}, \vec{v}}  (\vec{\omega}, T)$}  
describes the extent to which the applied control ``filters'' the effect of 
$O_{v_1} \cdots \,O_{v_\alpha} \otimes B_{u_1} (\omega_1) \cdots \,B_{u_\alpha} (\omega_\alpha)$ 
in $\Omega_\alpha(T)$.

In practice, one may have access to quantities that depend on the reduced (or ensemble-averaged, for 
classical noise) dynamics of the system alone.  The time-evolved reduced state, 
$\rho_S(T) \equiv \sum_{\ell , \ell'} \rho_{\ell \ell'} (T) |\ell \rangle\langle \ell'|$ with respect to an orthonormal 
basis on $S$, may be computed by averaging over the bath variables. Choosing a basis where the target 
gate $Q$ is diagonal, with $Q\ket{\ell}= q_\ell \ket{\ell}$, and assuming as usual that the initial state has the 
factorized form $\rho_{SB}(0) \equiv \rho_S(0) \otimes \rho_B$, we thus need 
\begin{eqnarray*}
&& \rho_{\ell \ell'} (T) =  q_\ell^\ast  q_{\ell'} \bra{\ell}  \textrm{Tr}_B 
[ \tilde{U}_e(T) \rho_{S} (0) \otimes \rho_B  \tilde{U}_e (T)^\dagger]  \ket{\ell'} \\
&& =  q_\ell^\ast q_{\ell'} \hspace*{-2mm} \sum_{m,m'} \rho_{m,m'}(0) \textrm{Tr}_B 
[ \bra{\ell} \tilde{U}_e(T) \ket{m} \rho_B  \bra{m'} \tilde{U}_e (T)^\dagger   
\ket{\ell'} ]. 
\end{eqnarray*}
Formally, we may Taylor-expand the error propagator given in Eq. (\ref{Magnus}), say, 
$\tilde{U}_e(T) =\sum_{r=0}^\infty ( {-i}\, \Omega_e(T))^r /r!$.  Then 
the trace over $B$ in the above expression is given by an infinite sum of terms each of which, 
fixing  $r$ and $r'$ (in the corresponding expansion of $\tilde{U}_e(T)^\dagger$), and 
taking advantage of Eq. (\ref{GFF}), may be recognized to have the following {\em symbolic} structure: 
\begin{widetext}
\begin{eqnarray*}
&& \int \hspace*{-1mm} {\cal D}\vec{\omega} \, 
G^{(\alpha_1)} \hspace*{-0mm}\cdots G^{(\alpha_r)}  G^{\ast (\alpha'_1)} \hspace*{-0mm}\cdots 
G^{\ast (\alpha'_{r'})} 
\bra{\ell} \textrm{Tr}_B [O_{[\alpha_1]} \hspace*{-0mm}\cdots  O_{[\alpha_r]} \otimes 
B_{[\alpha_1]} \hspace*{-0mm}\cdots B_{[\alpha_r]}  \ket{m} \hspace*{-0.5mm} \rho_B \hspace*{-.5mm} \bra{m'}     
O_{[\alpha'_1]} \hspace*{-0mm}\cdots  O_{[\alpha'_{r'}]} \otimes 
B_{[\alpha'_1]} \hspace*{-0mm}\cdots B_{[\alpha'_{r'}]} ] \ket{\ell'} = \nonumber \\
&& 
\hspace*{7mm}\bra{\ell} O_{[\alpha_1]} \hspace*{-0mm}\cdots  O_{[\alpha_r]} \ket{m}\bra{m'} 
O_{[\alpha'_1]} \hspace*{-0mm}\cdots  O_{[\alpha'_{r'}]}  \ket{\ell'}
\int \hspace*{-1mm} {\cal D}\vec{\omega}\, 
\underbrace{G^{(\alpha_1)}_{ \mbox{}_{{}_{}} } \hspace*{-0mm}\cdots G^{(\alpha_r)}  G^{\ast (\alpha'_1)} \hspace*{-0mm}\cdots 
G^{\ast (\alpha'_{r'})} }_{\textrm{filtering properties}}  \,
 \underbrace{ \textrm{Tr}_B [ \rho_B B_{[\alpha_1]} \hspace*{-0mm}\cdots B_{[\alpha_r]} 
B_{[\alpha'_1]} \hspace*{-0mm}\cdots B_{[\alpha'_{r'}]} ] }_{\textrm{noise properties}} .
\label{spectrum}
\end{eqnarray*}
\end{widetext}
Here, the additional indexes and arguments of each generalized FF $G^{(\alpha_j)} \equiv  
G^{(\alpha_j)}_{\vec{u}_{\alpha_j} \vec{v}_{\alpha_j}} ({\omega_1},\ldots \omega_{\alpha_j}, T)$ 
have been left implicit, we have used shorthand notations like $O_{[\alpha_j]}$ ($B_{[\alpha'_j]}$) to 
denote strings of system (bath) operators of length $\alpha_j$ ($\alpha'_j$), 
and $\int \hspace*{-1mm}{\mathcal D}\vec{\omega}$ denotes a multidimensional integral over all 
the relevant frequency variables.  
The main point is that in each term it is possible to clearly identify a filtering action 
on quantities that depend solely on ``high-order moments'' of the noise in frequency space,  
taken (in the quantum case) with respect to the initial state of $B$. 
In fact, such bath expectations may be related to the (high-order) noise power spectra given by 
the Fourier transform of the appropriate cumulant.  Following \cite{Mendel,supplement}, we 
may define the $k$-th order noise power spectrum: 
\beqy
\nonumber S^{(k)}_{\vec{u}} (\vec{\omega}) 
\nonumber &=& \int_{-\infty}^\infty \, d^k \vec{t}    \, e^{-i \vec{\omega} \cdot \vec{t} }\,
{\mathcal C} (B_{u_1} (t_1) \cdots B_{u_k} (t_k))\\
\label{powerspectra} &=& {\mathcal C}(B_{u_1} (\omega_1) \cdots B_{u_k} (\omega_k)),
\eneqy
where ${\mathcal C}(B_{u_1} (t_1) \cdots B_{u_k} (t_k))$ is the $k$-th order (generalized) 
cumulant with respect to $\rho_B$, that is, computed from the appropriate moment expansion by taking expectations 
with respect to $\rho_B$ and retaining operator ordering \cite{Kubo,analyticity}. 

In this way, one may make contact with the standard FF formalism 
\cite{Uhrig:07,PhysRevB.77.174509,WDD,Todd}.  The simplest 
setting is provided by ``bang-bang" DD on a single qubit exposed to either classical 
zero-mean, stationary Gaussian phase noise or spin-boson dephasing from a thermal bath.  
Without loss of generality, we can let $\tilde{H}_e(t) = y(t) \sigma_z \otimes B(t)$, with the 
``switching'' function 
$y(t)=\pm 1$ being determined by the applied pulse sequence $p$. The Gaussian 
statistics allows for the infinite expansion determining the coherence matrix element 
$\rho_{01}(T)\equiv \rho_{01}(0)e^{-\chi_p(T)}$ to be exactly re-summed \cite{supplement}, 
with $\chi_p (T) = 2 \int_{-\infty}^\infty d\omega/(2\pi) \, G^{(1)}(\omega, T) G^{(1)}(-\omega,T) 
S^{(2)}(\omega,-\omega)$, where $G^{(1)}(\omega, T) =  \int_{0}^T dt  \,y(t) e^{i \omega t}$  
and, thanks to stationarity, $S^{(2)}(\omega, \omega')= 2\pi \delta(\omega+\omega') S(\omega)$. 
Thus, the ``traditional'' FF is recovered as 
$F_p(\omega,T) \equiv  \omega^2 G^{(1)}(\omega, T) G^{(1)}(-\omega, T)$.  By construction, 
if the FF is small around some frequency $\omega_0$, the contribution of the noise at that 
frequency is suppressed. This may be formalized by considering the Taylor expansion of 
the FF around $\omega_0$. For DD, one is interested in $\omega_0=0$, yielding 
$F_p(\omega,T) \propto (\omega T)^{2(\delta +1)}$ \cite{Uhrig:07,WDD}.  Thus, if a 
{\em filtering order} (FO) $\phi$ is defined by the frequency scaling of 
$G^{(1)}$ as $\vert \omega \vert \rightarrow 0$, 
CO and FO {\em coincide} in this exactly solvable model.

{\em Fundamental filter functions and filtering order.--} Away from special scenarios where a single FF 
fully characterizes the cancellation and filtering properties of the applied control, evaluating 
arbitrary generalized FFs $G^{(\alpha)}_{\vec{u}, \vec{v}}(\vec{\omega},T)$ may seem beyond reach 
in general.  Our key insight is to realize that, despite their complexity, generalized 
FFs share a {\em common structure}, determined by a still infinite but easily computable FF set.  
Let {\em FFFs of order $\alpha$} be defined as follows:
\begin{eqnarray}
\label{FFF}
& F^{(\alpha)}_{\vec{u} \vec{v}} (\vec{\omega},T) 
\equiv {(-i)^\alpha} \int_{V_\alpha} d^{\alpha} \vec{t} \,
\,\prod_{j=1}^\alpha \left( y_{u_j v_j} (t_j) e^{i \omega_j t_j}\right),
\end{eqnarray}
where $V_\alpha$ is the integration volume previously defined.
Let us also introduce the notation 
$s_k \equiv \sum_{r=1}^k \alpha_{r}$ and  $\vec{u}_{[s_{k-1},s_k]} \equiv (u_{s_{k-1}+1},\cdots,u_{s_{k}})$.
Our main result is then the following: 

\vspace*{2mm}

\noindent
{\bf Theorem 1.} {\em Arbitrary generalized FFs of order $\alpha$, $\alpha =1, \ldots, \infty$, 
may be exactly represented in terms of FFFs of order at most $\alpha$: }
\begin{eqnarray}
\label{GFFtoFFF}
&& { -i} G^{(\alpha)}_{\vec{u} \vec{v}} (\vec{\omega}, T) = F^{(\alpha)}_{\vec{u} \vec{v}} (\vec{\omega}, T) \\ 
\nonumber 
&&  - \sum_{j=2}^{\alpha}  \frac{(-1)^j}{j} \hspace*{-2mm}
\sum_{\substack{ \sum_{r=1}^j \hspace*{-.4mm}\alpha_r = \alpha \\ \{\alpha_r >0\} }} \, 
 \prod_{k=1}^{j} \, {F^{(\alpha_k)}_{\vec{u}_{[s_{k-1},s_k]} \vec{v}_{[s_{k-1},s_k]}}
(\vec{\omega}_{[s_{k-1},s_k]},T)}. 
\end{eqnarray}

\noindent
{\bf Proof.} By using the Dyson series expansion, the error propagator $\tilde{U}_e(T)$ in Eq. (\ref{Magnus}) 
is given by $ \tilde{U}_e(T) = {\mathbb I} + \sum_{\alpha=1}^\infty {\mathcal D}_\alpha(T)$, where the 
$\alpha$-th Dyson term is a time-ordered integral over $V_\alpha$ 
of a function of $\alpha$ products of $\tilde{H}_e (t_j)$.  In frequency space, the latter reads 
${\mathcal D}_\alpha(T) = \sum_{\vec{u},\vec{v}}\int \frac{d^\alpha \vec{\omega}} 
{(2\pi)^\alpha} \, F^{(\alpha)}_{ \vec{u} \vec{v}} (\vec{\omega}, T)   
O_{v_1}  \cdots \,O_{v_\alpha}  \otimes B_{u_1} (\omega_1) \cdots \,B_{u_\alpha} (\omega_\alpha)$.
 One then makes use of the well-known expression relating Dyson and Magnus terms of a given order 
\cite{Blanes:08}, namely,
$ \Omega_1={\mathcal D}_1$, $\Omega_\alpha = {\mathcal D}_\alpha - \sum_{j=2}^\alpha ((-1)^j/j) 
\sum_{\alpha_1 + \ldots \alpha_j = \alpha} ({\mathcal D}_{\alpha_1}\cdots {\mathcal D}_{\alpha_j}),$
for $\alpha \geq 2$.  By also expressing $\Omega_{\alpha}(T)$ in frequency space 
using Eq. (\ref{GFF}), the desired result follows by 
equating terms filtering the same term $O_{v_1}\cdots O_{v_\alpha} \otimes B_{u_1} (\omega_1)
\cdots B_{u_\alpha}(\omega_\alpha)$.
\hfill$\Box$
 
\vspace*{1mm}

Thanks to the above theorem, arbitrary generalized FFs are explicitly, {\em non-recursively} 
computable from the appropriate FFFs. In fact, FFFs suffice to fully characterize the 
cancellation and filtering capabilities that 
a control protocol can guarantee {\em under minimal assumptions on the noise model}. 
We will say that $G^{(\alpha)}_{\vec{u}\vec{v}} (\vec{\omega},T)$ has {\em generalized FO 
$\Phi^{(\alpha)}_{\vec{u}\vec{v}}$} and {\em generalized CO} $\Delta^{(\alpha)}_{\vec{u}\vec{u}}$ 
(around $\vec{\omega} = \vec{\omega}_0$ and $T=0$) if 
$G^{(\alpha)}_{\vec{u}\vec{v}} (\vec{\omega},T) \sim \mathcal{O} 
(m^{{\Phi^{(\alpha)}_{\vec{u}\vec{v}}}} (\vec{\omega}-\vec{\omega}_0) \,
T^{\Delta^{(\alpha)}_{\vec{u}\vec{v}} +1}  )$, where 
$m^{d} (\vec{x})$ denotes a degree-$d$ monomial in the components 
of $\vec{x}$.  One can similarly define {\em fundamental} FO and CO for each of the FFFs, say, 
$\phi_{\vec{u}\vec{v}}^{(\alpha)}$ and $\delta_{\vec{u}\vec{v}}^{(\alpha)}$, respectively. 
However, the contribution of operators resulting from $H_{e}(t)$ is 
filtered only by products of {\em relevant} generalized or fundamental FFs, 
for which  $O_{v_1} \cdots O_{v_{\alpha}} \neq {\mathbb I}_S$ 
in {\em at least one factor} of such products, whereas all others correspond to pure-bath evolution 
$H_{B}^\textrm{eff}$ that is ``irrelevant'' for $S$. For each Magnus order $\alpha$, we denote the relevant set 
by ${\mathcal R}_\alpha$.  

\vspace*{2mm}

\noindent 
{\bf Definition.} {\em The generalized and fundamental CO $(\Delta,\delta)$ 
of a control protocol are given by the minimum of the corresponding quantity over 
the set of all the relevant FFs:}
\begin{align}
\Delta = \min_{ \mathcal{R}_\alpha, \forall \alpha }  \{\Delta_{\vec{u} \vec{v}}^{(\alpha)}\}, &\quad
\delta = \min_{ \mathcal{R}_\alpha, \forall \alpha }  \{\delta_{\vec{u} \vec{v}}^{(\alpha)}\}. 
\label{CO}
\end{align}
{\em Similarly, the generalized and fundamental FO up to level $\kappa$, 
$(\Phi^{[\kappa]}, \phi^{[\kappa]})$, $\kappa=1,\ldots,\infty$, 
are obtained by minimizing over the set of all the relevant  
FFs for $\alpha \leq \kappa$: }
\begin{align}
\Phi^{[\kappa]} = \min_{ \mathcal{R}_\alpha ,\alpha \leq \kappa } \{\Phi_{\vec{u} \vec{v}}^{(\alpha)} \},& \quad
\phi^{[\kappa]} = \min_{ \mathcal{R}_\alpha ,\alpha \leq \kappa } \{\phi_{\vec{u} \vec{v}}^{(\alpha)}\}. & 
\label{FO}
\end{align}

\vspace*{1mm}

\noindent 
When specific knowledge is available on the error model, we may 
further restrict the above minimizations to a smaller set of relevant FFs, yielding {\em effective 
FOs or COs} that may be higher than in Eqs. (\ref{CO})-(\ref{FO}).  For zero-mean Gaussian 
dephasing as previously discussed, for instance, 
since only {\em even} powers of $G^{(1)}(\omega,T)$ contribute to 
the reduced dynamics, it follows that $\Phi^{[\infty]}_{\text{eff}} = 2 \Phi^{[\infty]}$, 
in line with the standard analysis based on $F_p (\omega, T)/\omega^2$. Thus, the quantities defined 
in Eqs. (\ref{CO})-(\ref{FO}) yield {\em lower bounds} in general.  The following result holds: 

\vspace*{2mm}

\noindent
{\bf Theorem 2.} {\em The generalized and fundamental FO $(\Phi^{[\kappa]}, \phi^{[\kappa]})$ 
and CO $(\Delta,\delta)$ obey the following relationships: }
\begin{equation}
\Phi^{[\kappa]} = \phi^{[\kappa]}, \; \; \kappa = 1,\ldots, \infty ; \quad \Delta = \delta; 
\quad \phi^ {[\infty]} \leq \delta. 
\label{FOCO}
\end{equation}

\noindent
{\bf Proof.} By expressing $G^{(\alpha)}_{ \vec{u}\vec{v} }$ in terms of FFFs, 
Eq. (\ref{GFFtoFFF}), it follows that $\Phi^{(\alpha)}_{\vec{u}\vec{v}}  
\geq  \min_{\mathcal{R}_{\alpha' \leq \alpha}} \{\phi_{\vec{u} \vec{v}}^{(\alpha')}\}$,
for each fixed $\alpha \geq 1$ and $(\vec{u},\vec{v})$.  
Hence, minimizing the left hand-side yields 
$\Phi^{[\kappa]} =\min_{\mathcal{R}_{\alpha\leq \kappa}}  \Phi^{(\alpha)}_{\vec{u}\vec{v}}  
\geq  \min_{\mathcal{R}_{\alpha'\leq \kappa}} \{\phi_{\vec{u} \vec{v}}^{(\alpha')}\} =\phi^{[\kappa]}$.
However, by inverting the relationship between Magnus and Dyson terms \cite{Blanes:08}, 
one may also similarly express each 
$F^{(\alpha)}_{ \vec{u}\vec{v} }$ in terms of generalized FFs. Thus, we also have $\Phi^{[\kappa]} \leq \phi^{[\kappa]}$, whereby the first equality in Eq. (\ref{FOCO}) follows.  By a similar reasoning, 
minimizing over all ${\cal R_\alpha}$, the equality $\Delta = \delta$ also follows.

To establish the third inequality in Eq. (\ref{FOCO}), note that each $F^{(\alpha)}_{ \vec{u}\vec{v} }$ 
has dimensions $[T^\alpha]$. By definition of the associated FO and CO, it must also be that 
$ T^\alpha \sim T^{-\Phi^{(\alpha)}_{\vec{u}\vec{v}}    +\Delta^{(\alpha)}_{\vec{u}\vec{v}}   +1 }$. 
Since by definition $\alpha \geq 1$, it thus follows that 
$ \Phi^{(\alpha)}_{\vec{u}\vec{v}} \leq \Delta^{(\alpha)}_{\vec{u}\vec{v}}$.  The final step is to 
minimize over ${\cal R}_\alpha$, for all $\alpha$. \hfill$\Box$ 

\vspace*{1mm}

{\em Discussion.--} 
In order to gain insight into the general results described above, 
we first consider {\em single-axis} control protocols -- in particular, the ideal single-qubit DD 
setting as before, but now in the presence of arbitrary, {\em non-Gaussian} dephasing.  In this case,  
it is well known that CO $\delta$ over an evolution time $T$ may be achieved by 
using either (single-axis) concatenated DD \cite{KhodjastehLidar:04} or Uhrig DD 
\cite{Uhrig:07} -- say, CDD$_\delta$ and UDD$_\delta$, respectively, with the latter protocol 
being optimal in terms of the required number of control pulses, also equal to $\delta$.  For 
CDD$_\delta$, thanks to the symmetry properties that the concatenated structure grants to 
both the FFFs of lower level and the control functions $y_{uv}(t)$, one may easily 
prove the following \cite{supplement}: 

\vspace*{2mm}

\noindent
{\bf Proposition 1.} 
{\em Arbitrarily high FO may be achieved for ideal 
single-axis DD via concatenation:} $\phi^{[\infty]} = \delta$ for CDD$_\delta.$

\vspace*{1mm}

\noindent 
UDD protocols behave very differently.   Recalling that 
no even-order FFF is relevant since $\sigma^2_z={\mathbb I}_S$,
we have explicitly computed the first odd FFFs for COs $\delta= 1,\ldots, 8$.
We find that $\phi^{(1)}_{z}  =  \delta$,  $\phi^{(3)}_{zzz} = \delta-2$, and 
\beqynn 
\phi^{(5)}_{zzzzz}= \begin{cases} 
\delta-2 , \; \delta \in [3,4] ,\\ 
\delta-4 , \; \delta \in [5,8] , 
\end{cases} \hspace*{-2mm}
\phi^{(7)}_{zzzzzzz}= \begin{cases} 
\delta-2 ,\; \delta \in [3,6] , \\ 
\delta-6 , \; \delta \in [7,8] , 
\end{cases} 
\eneqynn
implying that the FO $\phi^{[\infty]} \leq 1$ or $2$ up to CO $\delta = 8$, depending of whether 
$\delta$ is odd or even. We conjecture that this holds for $\delta >8$.
% \cite{LongFFF}. 
Thus, while both UDD and CDD ensure arbitrary CO, only the latter guarantees an arbitrary FO as well. 
 
\begin{figure}[t]
	\centering
         \includegraphics[width=.93\columnwidth]{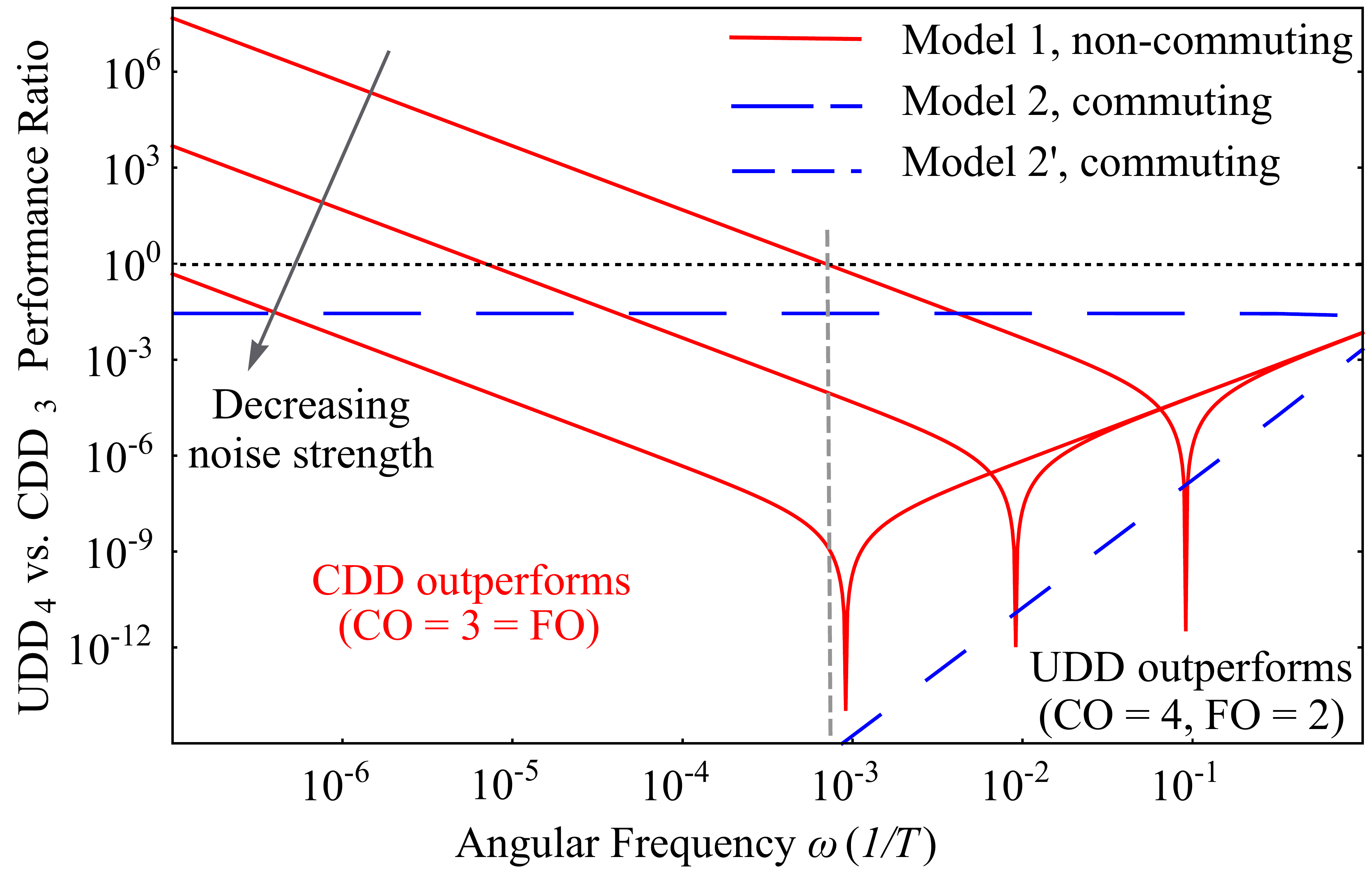}
	\vspace*{-2mm}
	\caption{(Color online) Performance of CDD$_3$ and UDD$_4$ for ``commuting'' (dashed lines) vs. ``non-commuting'' (solid lines) single-qubit dephasing models (see text).  A performance ratio $\Vert H_{SB}^{\textrm{eff}}(T)\Vert_{\textrm{UDD}_4}/\Vert H_{SB}^{\textrm{eff}}(T)\Vert_{\textrm{CDD}_3}<1$ (black dotted line) implies that UDD$_4$ outperforms CDD$_3$. Only the latter model is sensitive to the difference in FO, which manifests in the change in slope of the solid lines in the low-frequency regime ($\omega \lesssim 10^{-3}$ ). 
	Parameters are $T= 1$ and $g=9/40,9/400,9/4000$ in appropriate units, 
	such that $\Vert \tilde{H}_{e,1 (2)} (t) T\Vert < 1$ for all $t\in [0, T]$ and convergence is ensured.}
	\label{compa}
\end{figure}

The implications of this difference may be appreciated by contrasting two dephasing toy models: 
in model 1, $B$ consists of a qubit, such that 
$\tilde{H}_{e,1} (t) = g y(t) \sigma_z^{(1)} \otimes   [\cos (\omega t ) \sigma_z^{(2)} + 
\sin (\omega t) \sigma_y^{(2)}]$, whereas model 2 may be thought of as a classical version, with, say,  
$\tilde{H}_{e,2} (t) = g y(t) \cos (\omega t ) \sigma_z^{(1)}$, or 
$\tilde{H}'_{e,2} (t) = g y(t) \sin (\omega t ) \sigma_z^{(1)}$.
Here, $g$ is an overall coupling constant, and using a single frequency ``tone'' 
allows one to study the response to specific values of $\omega$.  A comparison between 
CDD$_3$ and UDD$_4$ is given in Fig. \ref{compa}, where in order to
avoid making assumptions about the initial state of $S$ and $B$ and the statistical properties of the noise, 
we choose a performance metric inspired by fault-tolerance analysis~\cite{Preskill2013}, 
$\Vert T  H_{SB}^{\textrm{eff}}(T)\Vert$, here computed via the Magnus expansion up to the third order. 
Due to its commuting nature, model 2 is (trivially) insensitive to the difference in FO.
Thus, UDD$_4$ always outperforms CDD$_3$, 
since the higher CO is all that matters when a single FF is important. 
In contrast, since high-order FFs contribute to $\Vert T  H_{SB}^{\textrm{eff}}(T)\Vert$ for model 1, 
the difference in FO translates into the existence of a low-frequency 
regime in which CDD$_3$ outperforms UDD$_4$, despite its lower CO. Since 
the effect of the higher-order Magnus terms is reduced as $g$ decreases, this frequency range 
is, correspondingly, reduced.  Despite its simplicity, this example thus clearly indicates that the 
FO can be the key property when the task is to remove noise that is stronger in a particular frequency 
range.

For noise models and/or control objectives requiring {\em multi-axis} protocols, 
aiming to arbitrarily high-order filtering is too stringent a requirement in general, even for a single qubit:

\vspace*{2mm}

\noindent
{\bf Proposition 2.} 
A protocol which does not achieve perfect cancellation of an arbitrary quasi-static noise model 
(i.e., one where $\{{B}_u (t)\}$ are constant), has $\phi^{[\infty]} =0$.
\vspace*{1mm}

\noindent
Existing DD/DCG control sequences subject to a minimum switching-time constraint
fall in this category (see \cite{supplement} for a sketch of the proof). 
While we conjecture that the above ``no-go'' result may hold more generally for control protocols subject 
to physical constraints. 
%\cite{LongFFF}, 
the practical implication is that, in general, it is only meaningful to demand a high 
FO for a {\em subset} of the GFFs -- e.g., those responsible for filtering the dominant noise contributions 
over a frequency range, that is, $\Phi^{[\kappa]}$ large for some $\kappa<\infty$. 
Remarkably, within the validity of a first-order fidelity approximation, 
the distinction between CO and such ``truncated'' ($\kappa=2$) FO has been already observed 
for well-known composite-pulse protocols \cite{KenReview}: e.g., so-called Solavay-Kitaev SK1 
and Wimperis BB1 sequences for amplitude errors have same FO ($=1$) despite being first- 
and second-order in the Magnus sense, respectively \cite{Chingiz,Soare2014}.  Likewise, 
Walsh-modulated logic gates with desired noise-filtering features against dephasing noise 
have been recently implemented in trapped-ion experiments  \cite{Soare2014}.   

Beyond the single-qubit setting, the FFF formalism presented here has proved instrumental to 
design DD protocols for multi-qubit dephasing that are more resource-efficient 
than existing ones \cite{multiDD}. While further investigation is needed to fully elucidate the 
significance of the noise-filtering perspective for dynamical quantum error control, 
we believe that our results provide a concrete starting point to meet this challenge. 

We thank Paola Cappellaro, Alexandre Cooper-Roy, and Kaveh Khodjasteh for useful input, 
and Michael Biercuk for constructive feedback and a critical reading of the manuscript. 
Partially supported from the US ARO (contract No. W911NF-11-1-0068)
and the Constance and Walter Burke Special Projects Fund in Quantum Information Science.

\bibliographystyle{apsrev}
\bibliography{bib}

%%%%%%%%%%%%%%%%%%%%%%%%%%%%%%%%%%%%%%%%%%%%%%%%%%

\onecolumngrid 

\section*{Supplementary Information}

\setcounter{figure}{0}
\setcounter{equation}{0}
\setcounter{section}{0}

\subsection{The simplest example: Single-qubit single-axis control} 
\label{simplest}

Consider a single-qubit model system, in which the internal Hamiltonian $H_S$ may be taken 
to be zero (in a suitable frame) and noise acts only along a known axis, chosen here to be the $z$ 
axis.  Assume that control is implemented in terms of sequences of perfect instantaneous (bang-bang) 
$\pi$ pulses, and that the control objective is to achieve DD ($Q ={\mathbb I}_S$).  
In the interaction picture with respect to the bath Hamiltonian, $H_B$, we may write
\beq
\label{depha}
\tilde{H}_e(t) = y(t) \,\sigma_z \otimes B(t), 
\eneq
where for simplicity we have denoted the control switching function $y_{zz}(t) \equiv y(t)$ and, 
likewise, $B_z(t)\equiv B(t)$ [cf. Eq. (1) in the 
main text].  The error propagator induced by such Hamiltonian after an evolution time $T$ is given by 
\beq
\tilde{U}_e (T) = \mathcal{T}_+ e^{ -i \int_0^T \tilde{H}_e(t) dt} = e^{-i T (H_{SB}^\textrm{eff}(T) + H_{B}^\textrm{eff}(T))} ,
\eneq 
where $\mathcal{T}_+$ denotes the time-ordering operator. 
By taking advantage of the fact that $(\sigma_z)^{2m}=
\mathbb{I}_S$, we may express the relevant error-action contributions in terms of the generalized FFs and the 
Fourier transforms of $B(t)$ defined in the main text [Eq. (3)]: 
\begin{align}
\label{hsb} T H_{SB}^\textrm{eff} (T) &= {\sigma}_z \otimes \sum_{\alpha:\text{odd}} G^{(\alpha)} (\vec{\omega},T) B(\omega_1) \cdots B(\omega_\alpha) , \\
\label{hb}  T H_{B}^\textrm{eff} (T) & = \sum_{\alpha:\text{even}} G^{(\alpha)} (\vec{\omega},T) B(\omega_1) \cdots B(\omega_\alpha). 
\end{align}

It is worth highlighting that since we are working in the interaction picture with respect to $H_B$, the error Hamiltonian $H_e(t)$ is explicitly time-dependent even before moving to the control toggling-frame, and hence the traditional definition of decoupling order, $\delta$, as determined by the lowest-order Magnus term which acts solely on the bath, does not apply. In fact, this is evident from a direct calculation in the presence 
of spin-boson dephasing in the interaction picture, see for example \cite{Uhrig:07}: even the first Magnus term has a non-trivial action on the system, and yet decoherence suppression is still achieved. 
Notice however, that the main consequence of demanding the first $\delta$ Magnus terms to be pure-bath 
in the standard toggling-frame setting, and what is truly essential for achieving decoherence suppression, 
is that the norm of the residual system-bath interaction is sufficiently small: that is, we require 
$\vert \vert T H_{SB}^\textrm{eff}(T) \vert \vert \sim \mathcal{O} (T^{\delta+1})$ for order-$\delta$ DD, thus yielding an interaction-picture evolution
$U(T)= \tilde{U}_e(T) = \mathbb{I}_S \otimes U_B(T) + \mathcal{O} (T^{\delta+1})$, for some unitary pure-bath operator $U_B(T)$. 
Defining, as in the main text, the {\em cancellation order} (CO) in this way is natural when working in the interaction picture or, more generally, whenever one has a time-dependent classical or quantum noise source, thereby formally allowing for a unified treatment.  We also refer to Ng {\em et al.} \cite{ng-2011} for additional discussion of DD in the interaction picture. 

Back to the problem at hand, we wish to determine the qubit reduced density matrix at time $T$, 
$\rho_S(T)$, specified by matrix elements $\rho_{\ell m}(T)$ with respect to the $z$ basis. 
By assuming an initial factorized state, $\rho_{SB}(0) = \rho_S(0) \otimes \rho_B$ 
(this assumption can in fact be weakened, as we shall further discuss in a future work),
%\cite{LongFFF}), 
and using Eq. (\ref{hsb}) above, we have:
\begin{align}
\nonumber \rho_{\ell m} (T) &= \bra{\ell} \textrm{Tr}_B [\tilde{U}_e(\tau) \rho_S(0) \otimes \rho_B \tilde{U}_e^\dagger(T) ]  \ket{m}\\
\label{reduced}&= \rho_{\ell m}(0) \, \textrm{Tr}_B [ e^{ -i [(-1)^\ell - (-1)^m] \sum_{\alpha:\text{odd}} \int 
\frac{d^\alpha \vec{\omega}}{(2\pi)^\alpha}  G^{(\alpha)}(\vec{\omega},T) B(\omega_1) \cdots B (\omega_\alpha) } \rho_B ].
\end{align}
The partial trace in square brackets can be evaluated using the following generalized cumulant expansion \cite{Kubo}. By letting $\langle \cdot \rangle_q \equiv \text{Tr}_B [ \cdot \, \rho_B]$, the cumulant expansion of 
$\langle e^P \rangle_q$, where $P$ is a generic bounded linear operator, 
is given by
\beq
\label{cumulant}
\langle e^P \rangle_q = e^{\sum_{k=1}^\infty \frac{C^{(k)}(P^k)}{k!}}, 
\eneq
where $C^{(k)}(P^k)$ is the $k$-th cumulant of $P$, which is assumed to be linear in its arguments (i.e., 
$C^{(k)}(P_1^k + P_2^k) = C^{(k)}(P_1^k) + C^{(k)}(P_2^k)$).
By performing a symbolic expansion of the exponentials on both sides of the equality and associating $C^{(k)}(P^k)$ with terms where $P$ appears $k$ times, one finds that cumulants can be expressed in terms of high-order moments $\langle \cdot \rangle_q$, much in the same way as in classical statistical mechanics they are written in terms of standard high-order moments, $\langle \cdot \rangle_c $~\cite{Kubo}. 
Of course, the treatment is equally valid if one replaces the quantum bath with a classical bath. 
In that case, as noted in the main text, we may let $B(t) = \beta(t) \mathbb{I}_B$, with $\beta(t)$ being a stochastic process and $\langle \cdot \rangle_q$ being 
replaced by the corresponding classical expectation, $\langle \cdot \rangle_c$, 
Because of the infinite sum in the exponent of the right hand side of Eq~\eqref{cumulant}, writing a closed-form solution or getting useful insight is not always straightforward, but some cases of interest allow considerable simplifications. In particular, we consider 
two such scenarios here: (i) Gaussian (quantum or classical) dephasing noise; and (ii) arbitrary classical 
dephasing noise.

{\it Gaussian dephasing.--} A Gaussian process strongly constrains the high-order cumulants, allowing us to truncate the infinite sum and get an exact solution. Classically, a noise process $\beta(t)$ is said to be Gaussian if $C^{(k)}(b(t_1) \cdots b(t_k)) = 0$ for $k>2$, and of zero-mean if $C^{(k=1)}( b(t)) =0$~\cite{Kubo}. 
In order to generalize this feature to the quantum case, 
one needs to be able to define the cumulant $C^{(k)}(B(t_1) \cdots B(t_k))$ in such a way that the non-commuting character of $B(t)$ at different times is properly taken into account. This can be done in a straightforward fashion by considering 
$P= B(t_1) + \ldots + B(t_k)$ and using the above generalized cumulant expansion, \eq{cumulant}. 
By identifying terms where the operators $\{ B(t_i) \} $ appear in the {\it fixed order} within each cumulant, one can easily write any $C^{(k)}( B(t_1) \cdots B(t_k))$ in terms of $\langle \cdot \rangle_q$, and vice versa. For example, after the formal expansion of the exponentials in Eq.~\eqref{cumulant}, the ``order $P^3$'' equation yields $$\langle P^3 \rangle_q = C^{(3)} (P^3) + 3 C^{(2)} (P^2) C^{(1)} (P) + C^{(1)} (P) C^{(1)} (P) C^{(1)} (P),$$ and the linearity of the cumulants allows the identifications
\begin{align*}
\langle B(t_2) B(t_3) B(t_1)\rangle_q &= C^{(3)}(B(t_2)B(t_3)B(t_1)) + C^{(1)}(B(t_2))C^{(1)}(B(t_3))C^{(1)}(B(t_1))  \\ 
& \,\,\,\, + C^{(2)}(B(t_2)B(t_3)) C^{(1)}(B(t_1)) + C^{(2)}(B(t_2)B(t_1)) C^{(1)}(B(t_3)) \\
& \,\,\,\,  +  C^{(2)}(B(t_3) B(t_1)) C^{(1)}(B(t_2)), \\
\langle  B(t_3) B(t_1) B(t_2)\rangle_q &=  C^{(3)}(B(t_3)B(t_1)B(t_2)) +  C^{(1)}(B(t_3))C^{(1)}(B(t_1))C^{(1)}(B(t_2)) \\ 
&\,\,\,\, + C^{(2)}(B(t_3)B(t_1)) C^{(1)}(B(t_2)) + C^{(2)}(B(t_3)B(t_2)) C^{(1)}(B(t_1)) \\
&\,\,\,\,  + C^{(2)}(B(t_1) B(t_2)) C^{(1)}(B(t_3)) .
\end{align*}
As expected, when arbitrary operators $B(t)$s at different times commute, both expressions are the same and one recovers the same relations one has for a classical noise process~\cite{Kubo}. Since in the quantum case $\langle\cdot \rangle_q$ is a function of the initial state of the bath $\rho_B$, we say that {\em the pair $\{B(t) , \rho_B\} $ defines a quantum Gaussian noise process if $C^{(k)} (B(t_1)\cdots B(t_k)) =0 $ for $k>2$ and, furthermore, the process has zero-mean when $C^{(k=1)} (B(t)) =0$}.  The well-known linear spin-boson dephasing model falls in this category whenever the initial state of the bath is thermal, see e.g. \cite{Preskill2013,multiDD}. 

Thus, when the noise is Gaussian (classical or quantum) and has zero-mean, the expression for the reduced dynamics, Eq. (\ref{reduced}),  drastically simplifies and we get
\begin{align}
\nonumber \rho_{01} (T) &= \rho_{0 1} (0) \, e^{\frac{1}{2}\langle \left( -2 i \int_{-\infty}^\infty \frac{d\omega}{2\pi} G^{(1)} (\omega,T ) B(\omega) \right)^2 \rangle }\\
&= \rho_{01} (0) \, e^{ - 2  \int_{-\infty}^\infty \frac{d\omega_1}{2\pi}  \int_{-\infty}^\infty \frac{d\omega_2}{2\pi}  G^{(1)}(\omega_1,T) G^{(1)} (\omega_2,T) \,\, \langle B(\omega_1) B(\omega_2)  \rangle},
\end{align}
with $\langle \cdot \rangle$ now denoting either $\langle \cdot \rangle_c$ or $\langle \cdot \rangle_q$. 
This yields the expression quoted in the main text for the decay rate $\chi_p(T)$ 
under the additional assumption that the noise is 
{\em stationary}: in such a case, by definition it must be $\langle B(t_1) B(t_2)  \rangle = \langle B(t) B(t+ t_2-t_1)  \rangle$ for all $t$, whereby it follows that 
\beq 
\langle B(\omega_1) B(\omega_2)  \rangle \equiv S^{(2)}(\omega_1,\omega_2)  = 2\pi \delta (\omega_1+\omega_2) S(\omega_2).
\eneq
As noted in the main text, the relevant FF $G^{(1)} (\omega,T )$ does not contribute directly in this case, but rather only through bilinear products $G^{(1)} (\omega_1,T ) G^{(1)} (\omega_2,T )$.  More general (e.g. non-zero mean) dephasing models may be similarly handled. 
%, see \cite{LongFFF} for further treatment.

{\it Arbitrary classical dephasing.--} When the noise is classical, i.e., $B(t) = \beta(t) \mathbb{I}_B$, the structure of the equations can also be simplified. In this case, because all the higher-order commutators in the Magnus expansion vanish, the relevant ensemble average is given by  
\begin{align}
\nonumber \langle \rho_{\ell m}(T) \rangle_c  &= \langle \, \bra{\ell}  \tilde{U}_e (T) \rho_S(0)  \tilde{U}_e^\dagger(T) \ket{m} \, \rangle_c \\
\nonumber &= \rho_{\ell m}(0) \, \langle e^{-i [(-1)^\ell - (-1)^m]  \int \frac{d {\omega}}{2\pi}  G^{(1)} ({\omega},T) \beta (\omega)} \rangle_c\\
\label{reducedclassical} &= \rho_{\ell m}(0)  \, e^{\sum_{k=1}^\infty \frac{ \{ i [ (-1)^m - (-1)^\ell ] \}^k}{k!}    \int \frac{d^k \vec{{\omega}}}{(2\pi)^k}  G^{(1)} ({\omega_1},T )\cdots G^{(1)} ({\omega_k}, T) \langle  \beta(\omega_1) \cdots \beta(\omega_k)\rangle_c}. 
\end{align}
Thus, similarly to the Gaussian case, all the effects of the control are condensed in a single FF, 
$G^{(1)} ({\omega},T ) \equiv F^{(1)} ({\omega},\tau)$.  However, products of arbitrary order may now 
be relevant in principle, depending on the non-Gaussian character of the noise statistics.  
It is worth highlighting that a similar simplification occurs in a dephasing (quantum) spin-boson noise model 
given a generic (non-Gaussian) initial bath state. In this case, higher order Magnus terms are also irrelevant, 
because they either vanish based on algebraic considerations or do not otherwise contribute to the reduced system dynamics \cite{multiDD}.

\subsection{Fundamental FFs as building blocks: Single-qubit multi-axis control } 

In the main text we showed that FFFs condense the filtering effect of a dynamical quantum error suppression protocol at the effective Hamiltonian and reduced dynamics level. It follows that this is also true for any functional of the propagator or the reduced density matrix of the system. To illustrate this claim and make contact with previous relevant work, we explicitly show how the ``effective'' generalized FFs introduced by 
Green {\em et al.} \cite{Todd} can be rewritten in terms of the generalized FFs $\{G^{(\alpha)}_{\vec{u}\vec{v}} (\vec{\omega},T )\} $ we introduced [Eqs. (3) and (6) in the main text]. 

In Refs. \cite{Todd}, the authors consider a (drift-less, $H_S=0$) single qubit exposed to a multi-axis classical noise process such that, using the notation of the main text, the time-dependent error Hamiltonian in the toggling frame with respect to the control may be written as 
\beq
\tilde{H}_e (t) = \sum_{u,v} y_{uv} (t) {\sigma}_v \beta_u(t), \quad u,v \in \{ x,y,z\}, 
\eneq 
where the stochastic process $\beta_u(t)$ describes classical noise in the $u$ spatial direction and 
$y_{uv}(t) \equiv [Y(t)]_{uv} = (Y_x(t), Y_y(t), Y_z(t))^T \in SO(3)$ is the relevant control matrix. 
If the propagator $U(T)$ resulting from the evolution induced by ${H}_{\text{ctrl}}(t)$ over a time $T$ would ideally execute a target gate $Q$, then, writing the total propagator in the presence of $H_e(t)$ as $U(T) = Q \tilde{U}_e(T)$ 
as in the main text, one can quantify the accuracy of a control operation subject to the noise process 
$\{\beta_u(t)\}$ via the operational (ensemble-averaged) fidelity
\beq 
\mathcal{F}_{\text{av}} (T) = \frac{1}{4} \, \langle | \textrm{Tr}( \tilde{U}_e(T)) |^2 \rangle .
\eneq
One may write the error propagator $\tilde{U}_e (T ) = \exp [- i \vec{a} (T ) \cdot \vec{\sigma} ]$, 
where $\vec{\sigma} = \{{\sigma}_x,{\sigma}_y,{\sigma}_z\}$ and $\vec{a} (T) \equiv  a(T) \hat{a}(T) = 
 \sum_{{\alpha}=1}^\infty [\vec{a}]_{\alpha} (T)$ is the ``error vector'' (in the terminology of \cite{Todd}), computable via the Magnus expansion. Specifically, each $[\vec{a}]_{\alpha} (T)$ originates from the ${\alpha}$-th order Magnus term and the component associated to the operator ${\sigma}_v$ is determined by 
\beq
\label{agral}
[a_v]_{\alpha}  (T) {\sigma}_v = \sum_{\substack{\vec{u}, \vec{v} \text{ such that} \\ {\sigma}_{v_1} \cdots \,
{\sigma}_{v_{\alpha}} = {\sigma}_v}} \int_{-\infty}^\infty \frac{d\omega_1}{2\pi} \cdots \int_{-\infty}^\infty \frac{d\omega_{\alpha}}{2\pi} \, G^{({\alpha})}_{\vec{u},\vec{v}} (\vec{\omega}, T) \, {\sigma}_{v_1} 
\cdots {\sigma}_{v_{\alpha}} \beta_{u_1} (\omega_1) \cdots \beta_{u_{\alpha}} (\omega_{\alpha}).
\eneq

By re-expressing the fidelity in terms of the length $a(T)$ of the error vector, $\mathcal{F}_{\text{av}} (T) 
= \frac{1}{2} [\langle \cos (2 a) \rangle +1 ]$,  Eq. (\ref{agral}) may be used as the starting point to obtain a fidelity expansion in terms of the generalized FFs and noise spectral density functions.  For instance, in the relevant case of dephasing noise, whereby only $\beta_{z}(t)$ is non vanishing, 
the elements of the control matrix are $y_{zx}(t),y_{zy}(t)$ and $y_{zz}(t)$.  By considering a Taylor expansion 
of the cosine term, the lowest order contributions to the fidelity are given by \cite{Todd}(b):
\beq
\label{fav}
 \mathcal{F}_{\text{av}} (T)   \simeq 1 - \sum_{v=x,y,z} \langle [a_v]_1^2 \rangle - \left\{ \sum_{v=x,y,z} \left(\langle [a_v]_2^2 \rangle  +2 \langle [a_v]_3 [a_v]_1 \rangle \right) - \frac{1}{3} \sum_{v,v'} \langle [a_v]_1^2 [a_{v'}]_1^2 \rangle \right\}.
\eneq
This expression can be written in terms of our generalized FFs $\{G^{(\alpha)}_{\vec{u}\vec{v}} (\vec{\omega},T )\}$, by using Eq. (\ref{agral}), e.g.:
\begin{align*}
[a_v]_1 (T) &= \int_{-\infty}^\infty \frac{d\omega}{2\pi}  G^{(1)}_{v}({\omega}, T) \beta_z (\omega), \quad 
v=x,y,z, \\
[a_x]_2 (T) &=  \int_{-\infty}^\infty \frac{d\omega_1}{2\pi} \int_{-\infty}^\infty \frac{d\omega_2}{2\pi} \,i \left( G^{(2)}_{y,z}({\omega_1,\omega_2}, T) - G^{(2)}_{z,y}({\omega_1,\omega_2}, T)\right) \beta_z (\omega_1) \beta_z (\omega_2),\\
[a_y]_2 (T) &=  \int_{-\infty}^\infty \frac{d\omega_1}{2\pi} \int_{-\infty}^\infty \frac{d\omega_2}{2\pi} \,i \left( G^{(2)}_{z,x}({\omega_1,\omega_2}, T) - G^{(2)}_{x,z}({\omega_1,\omega_2}, T)\right) \beta_z (\omega_1) \beta_z (\omega_2),\\
[a_z]_2 (T) &=  \int_{-\infty}^\infty \frac{d\omega_1}{2\pi} \int_{-\infty}^\infty \frac{d\omega_2}{2\pi} \,i \left( G^{(2)}_{x,y}({\omega_1,\omega_2}, T) - G^{(2)}_{y,x}({\omega_1,\omega_2}, T)\right) \beta_z (\omega_1) \beta_z (\omega_2),
\end{align*}
and so on. Under the additional assumptions that the noise is {\em stationary Gaussian and has zero mean}, 
the above expressions can be further simplified. Since  
\begin{align*} 
\langle \beta_z(\omega_1) \beta_z(\omega_2)\beta_z(\omega_3)\beta_z(\omega_4) \rangle &= \langle \beta_z(\omega_1) \beta_z(\omega_2) \rangle \langle \beta_z(\omega_3)\beta_z(\omega_4) \rangle +\langle \beta_z(\omega_1) \beta_z(\omega_3) \rangle \langle \beta_z(\omega_2)\beta_z(\omega_4) \rangle\\
& \,\,\,\,\, +\langle \beta_z(\omega_1) \beta_z(\omega_4) \rangle \langle \beta_z(\omega_2)\beta_z(\omega_3) \rangle, 
\end{align*}
the statistical properties of the noise are entirely captured by the power spectral density,  
$S_z(\omega) = \int_{-\infty}^\infty dt \,e^{-i \omega \tau} \langle \beta_z(t) \beta_z (t + \tau) \rangle$. 
One can then rewrite Eq. (\ref{fav}) in the frequency domain (see in particular Eq. (4) in the Supplementary 
Material of \cite{Todd}(b)):
\beq
 \mathcal{F}_{\text{av}} (T)   \simeq 1 - \frac{1}{4 \pi} \int_{0}^\infty \frac{d\omega}{\omega^2} S_z(\omega) F_1(\omega,T)  - \frac{1}{(4\pi)^2} \sum_p \int_0^\infty d\omega S_z(\omega)  \int_0^\infty \frac{d\omega'}{{\omega'}^2} S_z(\omega') F_{p,2} (\omega,\omega',T),
\eneq
where $F_1(\omega,T)$ and $F_{p,2} (\omega,\omega',\tau)$ are, in our context, {\em effective} generalized FFs computed through $\{ [a_v]_{\alpha}\}$, and the index $p$ generically refers to higher-order contributions. While we shall not proceed to further rewrite all their equations in terms of our $\{G^{(\alpha)}_{\vec{u}\vec{v}} (\vec{\omega},T )\}$ here, it should be clear at this point that $\{G^{(\alpha)}_{\vec{u}\vec{v}} (\vec{\omega},\tau)\}$, and by extension $\{F^{(\alpha)}_{\vec{u}\vec{v}} (\vec{\omega},T )\}$, appear as building blocks of the final expressions for the effective generalized FFs found in Ref.~\cite{Todd}.
For instance, it is easy to verify that 
\beq
\frac{F_1(\omega,T)}{\omega^2} = F^{(1)}_{x} (\omega,T)F^{(1)}_{x} (-\omega,T) +F^{(1)}_{z} (\omega,\tau)F^{(1)}_{z} (-\omega,T)+F^{(1)}_{y} (\omega,T)F^{(1)}_{y} (-\omega,T), 
\eneq
generalizing the correspondence $F_p(\omega, T)/\omega^2 = F^{(1)}(\omega,T)F^{(1)} (-\omega,T)$ already discussed for single-axis control under dephasing noise.

\subsection{Proof of Proposition 1}

We wish to show that, by assuming access to perfect instantaneous control pulses (say, along the $x$ axis), 
a FO $\phi^{[\infty]} =\delta$ is achieved by single-axis DD after $\delta$ levels of concatenation, 
for arbitrary dephasing noise on a qubit -- as claimed in the main text. 
 
The controlled evolution under a $\textrm{CDD}_k$ protocol, with CO equal to $k$,  
can be obtained via the recursion
\beq
\label{recu}
U_{k+1} (T_{k+1} = 2T_{k} ) = \sigma_x U_{k}(T_{k}) \sigma_x U_{k} (T_{k}) ,
\eneq
where  $U_0 (T_{0})$ denotes free evolution under a general dephasing Hamiltonian of the form 
$H = {\sigma}_z \otimes B + H_B$ over a time $T_0$. In the toggling frame associated to the control pulses and 
the interaction picture with respect to $H_B$, the joint dynamics is ruled by the Hamiltonian already given in Eq. (\ref{depha}) in the Supplement, 
$\tilde{H}_e (t)  = y(t) {\sigma}_z \otimes B (t)$, and CO $=\delta$ means that the resulting 
interaction-picture propagator obeys
\beq
U_{\delta}(T_\delta) = \mathbb{I}_S \otimes U_B(T_\delta) + \mathcal{O} (T_\delta^{\delta+1}) .
\eneq
We will show that this also achieves $\phi^{[\infty]} = \delta$. 
Recalling Eqs. (\ref{hsb})-(\ref{hb}) in the Supplement, the FFFs of interests are 
\beq
F^{(\alpha)} (\vec{\omega}, T_\delta) \equiv F^{(\alpha)}_{k=\delta},
\eneq
and in particular the ones with {\em odd} Magnus index $\alpha$, since those 
are associated with terms in $H_{SB}^{\textrm{eff} } (T_\delta)$. 

The starting point is to show that $F^{(\alpha)}_{k=1}$ has FO $\phi^{(\alpha)}_{k=1} \geq 1$ for all $\alpha$. 
Since $U_{1} (T_{1}) = \sigma_x U_{0}(T_{0}) \sigma_x U_{0} (T_{0})$, note that $y(t) = +1$ for $t \in [0,T_0]$ and $y(t) = -1$ for $t \in [T_0,2T_0]$, that is, the control switching function obeys the symmetry property 
$y(t) = - y(t + T_0)$. Exploiting this symmetry and dividing the length-$T_1$ integration intervals in subintervals of length $T_0$, one finds that the $\mathcal{O}(\omega^0)$ contribution to $F^{(\alpha)}_{k=1}$,  obtained by letting $\vec{\omega}=0$ in the expression for $F^{(\alpha)} (\vec{\omega}, T_1)$, is given by 
\begin{align*}
F^{(\alpha)}_{k=1} \vert_0  & = \int_0^{T_1} ds_1 \int_0^{s_1} ds_2 \cdots \int_0^{s_\alpha} 
ds_\alpha \, y(s_1) \cdots y(s_\alpha)  \\
&= [1 + (-1)^\alpha] \,F^{(\alpha)}_{k=0} \vert_0 + \sum_{a=1}^{\alpha-1} (-1)^{a} \,
F^{(a)}_{k=0} \vert_0 \, F^{(\alpha-a)}_{k=0} \vert_0 \\
&= [1 + (-1)^\alpha] \, \frac{T_0^\alpha }{\alpha!} + \sum_{a=1}^{\alpha-1} (-1)^{a} \,
\frac{T_0^a}{a!} \,\frac{T_0^{\alpha-a}}{(\alpha-a)!}\\
&= \frac{(T_0 - T_0)^\alpha}{\alpha!} =0.
\end{align*} 
Hence, $\phi^{(\alpha)}_{k=1} \geq 1$ for all $\alpha$. To proceed, we first observe that, as a consequence of the recursion in Eq.~\eqref{recu}, it also follows that $y(t) =- y(t + T_{\delta-1})$ for $t \in [0, T_{\delta-1}]$. Again, we exploit this symmetry by decomposing each FFF appropriately. Using the definition
\beq
I_{{q},{q}+1}^{l_1,l_2} (T_{\delta-1}) \equiv  \int_{q T_{\delta-1}}^{(q+1) T_{\delta-1}} ds_{l_1} \int_{q T_{\delta-1}}^{s_{l_1}} d{s_{l_{1}+1}} \cdots  \int_{q T_{\delta-1}}^{s_{l_2-1}} d s_{l_2} \, e^{i \sum_r \omega_r s_r} y(s_{l_1}) ... y (s_{l_2}) ,
\eneq
one finds that
\begin{align}
\nonumber F^{(\alpha)}_{k=\delta} &= \int_0^{T_\delta} ds_1 \int_0^{s_1} ds_2 \cdots \int_0^{s_{\alpha-1}} ds_\alpha \, e^{i \sum_r \omega_r s_r} y (s_1) ... y (s_j),  \\
\nonumber &=  I_{{0},1}^{1,\alpha} (T_{\delta-1}) + I_{{1},2}^{1,\alpha} (T_{\delta-1})  + \sum_{1 \leq a < \alpha } I_{{1},2}^{1,a} (T_{\delta-1})  \cdot I_{{0},{1}}^{a+1,\alpha} (T_{\delta-1})\\
\label{simcdd}  &= [1 + (-1)^\alpha e^{i \sum_r \omega_r T_{\delta-1}} ] \, I_{{0},1}^{1,\alpha} (T_{\delta-1}) + \sum_{1 \leq a < \alpha } I_{{1},2}^{1,a} (T_{\delta-1})  \cdot I_{{0},{1}}^{a+1,\alpha} (T_{\delta-1}).
\end{align}
Now, if $\alpha$ is odd, either $a$ or $\alpha- a$ are odd (but not both) and $[1 + (-1)^\alpha e^{i \sum_r \omega_r T_{\delta-1}}] \sim \mathcal{O}(m^1 (\vec{\omega}) T_{\delta-1})$, whereas if $\alpha$ is even, then 
$a$ and $\alpha-a$ must both be either odd or even and $[1 + (-1)^\alpha e^{i \sum_r \omega_r T_{\delta-1}}]
\sim \mathcal{O}(m^0 (\vec{\omega}))$. Notice that $I_{{q},q+1}^{1,a} (T_{\delta-1})$ and $I_{{q},q+1}^{a+1,\alpha} (T_{\delta-1})$ are proportional to $F^{(a)}(\vec{\omega},T_{\delta-1})$ and $F^{(\alpha-a)}(\vec{\omega},T_{\delta-1})$, respectively. Then, since the FO of the left hand side of Eq. (\ref{simcdd}) is lower-bounded by the FO of each term on the right hand side, the following recursion holds: 
\begin{align}
\phi^{(\alpha\textrm{:even})}_{k=\delta+1} & \geq \min [ \phi^{(\alpha\text{:even})}_{k=\delta}, \{ \phi^{(a)}_{k=\delta} + \phi^{(\alpha-a)}_{k=\delta} \}_{a <\alpha}]\\
\phi^{(\alpha\textrm{:odd})}_{k=\delta+1} & \geq \min [ \phi^{(\alpha\text{:odd})}_{k=\delta} + 1 , \{ \phi^{(a)}_{k=\delta} + \phi^{(\alpha-a)}_{k=\delta}\}_{\text{a:odd} <\alpha}]. 
\end{align}
Since we have already shown that $\phi^{(\alpha)}_{k=1} \geq 1$ for all $\alpha$, inductive reasoning implies 
that $\phi^{(\alpha\textrm{:even})}_{k=\delta} = 1$ and $ \phi^{(\alpha\textrm{:odd})}_{k=\delta} = \delta$, 
from which the desired result follows.  $\hfill\Box$

\subsection{Proof of Proposition 2}

We wish to show that if a protocol does not achieve perfect cancellation of an arbitrary 
quasi-static noise model, then it must have infinite-level FO $\phi^{[\infty]} =0$. We prove it by contradiction. 

Assume that the protocol in consideration, defined by the control matrix $\{ y_{uv}(t)\}$, does not achieve perfect cancellation of a quasi-static noise model but has $\phi^{[\infty]} >0$. Then one would have that $ F^{(\alpha)}_{\vec{u}\vec{v}} (\vec{\omega},T) $ is at least $\mathcal{O} (\omega^1)$, and thus it must be that (letting $\omega_j = 0\, \forall \,j$ in Eq. (5) of the main text)
\beq
\label{zeroth}
\int_{0}^T dt_1 \int_{0}^{t_1} dt_2 \cdots \int_{0}^{t_{\alpha-1}} dt_\alpha \,y_{u_1 v_1} (t_1) \cdots  
y_{u_\alpha v_\alpha} (t_\alpha) =0 ,
\eneq
for all relevant $\alpha$ and $\vec{u}, \vec{v}$, that is, such that $O_{v_1} \cdots O_{v_\alpha} \neq \mathbb{I}_S$. However, the integral in the left hand-side of Eq. (\ref{zeroth}) is precisely the integral appearing in every Magnus term when applying the control sequence to a quasi-static noise model, namely, one where every relevant bath operator $B_u(t)$ is constant. Thus, if a protocol has $\phi^{[\infty]} >0$ for an arbitrary (potentially unknown) noise model, then it must also be capable of {\em perfect cancellation} of any (again, potentially unknown) quasi-static noise model.  If so, the resulting (interaction-picture) propagator is exactly of the form $U(T) = \mathbb{I}_S \otimes U_B(T)$. Since by hypothesis cancellation is not, however, achieved perfectly, we have reached a contradiction. This implies that for any control protocol which does not achieve perfect cancellation of an arbitrary  quasi-static error model, there must exist (at least) a relevant $F^{(\alpha)}_{\vec{u},\vec{v}}$ such that $\phi^{(\alpha)}_{\vec{u}\vec{v}} = 0$, thus forcing 
$\phi^{[\infty]} = 0$. $\hfill\Box$

As mentioned in the main text, we conjecture that a stronger no-go results does in fact hold, namely, that the infinite-level FO $\phi^{[\infty]} =0$ for arbitrary open-loop control protocols subject to a physical minimum-switching time constraint.  Additional discussion and rigorous derivations will be presented elsewhere. 
%\cite{LongFFF,NoGo}.

\end{document}